\documentclass[aps,prb,reprint,amsmath,amssymb]{revtex4-2}
\usepackage{amsfonts}
\usepackage{dsfont}
\usepackage{cleveref}
\usepackage{booktabs}
\usepackage{graphicx}
\usepackage{url} 
\usepackage{color}

\begin{document}
	
	\title{Solving the Transient Dyson Equation with Quasilinear Complexity via Matrix Compression}
	
	\author{Baptiste LAMIC}
	\affiliation{Univ. Grenoble Alpes, CEA, IRIG-Pheliqs, F-38000 Grenoble, France}
	
	\date{\today}
	
	\begin{abstract}
We introduce a numerical strategy to efficiently solve the out-of-equilibrium Dyson equation in the transient regime. By discretizing the equation into a compact matrix form and applying state-of-the-art matrix compression techniques, we achieve significant improvements in computational efficiency which results in quasi-linear scaling of both time and space complexity with propagation time. This enables to compute accurate solutions even for systems with multiple and disparate time scales. We benchmark our solver by simulating a voltage-biased Josephson junction formed by a quantum dot connected to two superconducting leads.

	\end{abstract}
	
	\maketitle
	
	\section{Introduction}
	
The quantum field formalism provides an elegant framework for describing interacting and many-body quantum systems by providing 
tools for evaluating $n$-point correlation functions, thus assessing the statistical properties of observables in both equilibrium and non-equilibrium regimes \cite{Altland2010, Rammer2007}. This formalism underpins various numerical schemes, including quantum determinant Monte Carlo \cite{Profumo2015, Kozik2010}, functional renormalization group methods \cite{Metzner2012} and perturbation theories \cite{Rammer2007, Altland2010}, all of which are capable of directly accessing quantities in the thermodynamic limit. However, these methods often require the evaluation of multidimensional integral operators, which makes their numerical resolution computationally expensive. 

The Dyson equation $G = g + g \cdot \Sigma \cdot G$ is one example of such integral equation. It relates the full Green function $G$ which describes the two-points correlations in the system, to the free Green function $g$ representing the non-interacting system, and the self-energy $\Sigma$ which accounts for interaction effects. Solving this equation has proven to be challenging both at equilibrium \cite{Dong2020} and out-of-equilibrium \cite{Talarico2019}, yet some significant advancements were made by leveraging matrix compression techniques \cite{Kaye2021, KayeEquilibrium}. In this work, we refine the usage of these compressions methods to solve the out-of-equilibrium Dyson equation with unprecedented efficiency. 
	
Within the real-time formulation of the Keldysh out-of-equilibrium field theory, the Dyson equation can be expressed as follows
	\begin{equation}
		\small
		G(t,t') =  g(t,t') +  \iint_{\boldmath{-\infty}}^{\infty}  g(t,t_1)  \Sigma(t_1,t_2) G(t_2,t') \text{d}t_1 \text{d}t_2 	\label{eq:formal_dyson_equation}.
	\end{equation}	
We select the basis of the Keldysh space such that the kernels can be written as
	\begin{equation}
		 G \equiv 
		\begin{pmatrix}
			0 & G^\text{A} \\
			G^\text{R} & G^\text{K}
		\end{pmatrix}, \quad
		 \Sigma \equiv \begin{pmatrix}
			\Sigma^\text{K} & \Sigma^\text{R} \\
			\Sigma^\text{A} & 0
		\end{pmatrix}, 
		\label{eq:basis:def}
	\end{equation}
	where the superscripts $\text{R}$, $\text{A}$, and $\text{K}$ denote the retarded, advanced, and kinetic components, as detailed in \cite{Rammer2007}.  Any retarded or advanced kernel $F^\text{R/A}$ satisfy
	\begin{align}
		\small
		F^\text{R}(t,t') &= \theta(t-t')F^\text{R}(t,t'), \\
		F^\text{A}(t,t') &= \theta(t'-t)F^\text{A}(t,t').
	\end{align}
	Eq.\,\ref{eq:formal_dyson_equation} can be written as
	\begin{align}
		\small
		G^\text{R} &= g^\text{R} +  g^\text{R} \Sigma^\text{R} G^\text{R}, \label{eq:retarded_dyson_equation} \\
		G^\text{A} &= {G^\text{R}}^\dagger, \\
		G^\text{K} &= \left(\mathds{1}+G^\text{R}\Sigma^\text{R}\right)g^\text{K}\left(\mathds{1} + \Sigma^\text{A} G^\text{A}\right) + G^\text{R} \Sigma^\text{K} G^\text{A}, \label{eq:kinetic_dyson}
	\end{align}
	where all the integrations are implicit. We assume that the problem's orbital degrees of freedom have been discretized so that for any pair of kernels $(F,R)$, the product $F(t,t_1) R(t_1,t')$ satisfies:
	\begin{equation}
		\left[ F(t,t_1) R(t_1,t') \right]_{p,q} = \sum_{k} 	F_{p,k}(t,t_1) R_{k,q}(t_1,t').
	\end{equation}
	A natural starting point to solve \cref{eq:retarded_dyson_equation} is to discretize the time axis using a regular time grid $t_p = t_0 + p \delta_t$ for $0 \leq p \leq N$, where $\delta_t$ is the time step and $N+1$ is the number of points. Any kernel $F(t,t')$ is formally discretized into the block matrix $\mathbf{F}$. Each block is defined by
	 \begin{equation}
		\mathbf{F}_{i,j} \equiv F(t_i,t_j). 
	\end{equation}
	The traditional method stores each block explicitly, thus requiring $\mathcal{O}(N^2)$ space, and setting a lower bound on the computational complexity.  The retarded Dyson equation is usually solved using a time-stepping approach demanding $\mathcal{O}(N^3)$ operations as $N$ steps must be computed for $N$ values of $t'$ and each step requires summing over the full system history to evaluate the integral as shown in \cite{Talarico2019, Kaye2021}. Using \cref{eq:kinetic_dyson} to evaluate the  Keldysh component requires integrating over all the history of the system, starting at $-\infty$.  While there
	are methods to perform this integral \cite{NESSI, Kaye2021, Dong2020}, we impose in the following that the self-energy is turned on at $t = t_0$ and eventually turned off after $t_\text{end}$, \textit{i.e.} $\forall (t,t') \notin [t_0, t_\text{end}]^2: \Sigma(t,t')^{R/A/K} = 0$. This simplifies the discussion by ensuring that all integration domains lie within $[t_0, t_\text{end}]$.

As $N \sim \tau_{\text{long}} / \tau_{\text{short}}$, with $\tau_{\text{long}}$ and $\tau_{\text{short}}$ the longest and shortest system timescales, the traditional time-stepping method require $\mathcal{O}(\left(\tau_{\text{long}} / \tau_{\text{short}}\right)^3)$ operations. This unfavourable scaling has prevented the use of the Dyson equation in simulating quantum system with widely separated timescales. Alternative methods that sacrifice part of the quantum field elegance and flexibility for speed have emerged \cite{kloss2021tkwant, kloss2021tkwant, Tuovinen}, or solver specialized for specific form of the equation \cite{Ortega-Taberner2023, Venitucci2018, KayeEquilibrium}. However, recent works have reduced the required time and space to solve the Dyson equation  to $\mathcal{O}(N^2 \log(N))$ and $\mathcal{O}(N \log(N))$, respectively, by using matrix compression techniques to perform fast integral evaluation and compress the kernel representation \cite{Kaye2021}. Here, we take a step beyond  by proposing a new discretization of the Dyson equation in time domain that enables a more efficient use of matrices compression method. As a result, we reduce the time complexity from a quadratic scaling to the quasilinear scaling $\mathcal{O}(N \log(N))$ without further assumptions. 
	
\section{Algorithm}
We discretize the kernel products using the Nyström method with quadrature weights that remain constant away from the integration domain edges, a condition satisfied by Gregory quadratures \cite{Fornberg2019}. To avoid overwhelming the reader with indices, we propose here a compact formulation of the algorithm and we restrict the presentation to a low-order discretization using trapezoidal quadrature. An expanded version is presented in the supplemental materials \cite{Supplement}. Thus, we approximate the kernel product $A^\text{R} B^\text{R}$ as
\begin{equation}
	 (A^\text{R} B^\text{R})(t_p,t_q) \approx 	\mathbf{A^\text{R}}  * \mathbf{B^\text{R}} ,
\end{equation}
where we defined the discretized Kernel product as
		\begin{equation}
		\mathbf{A^\text{R}}  *\mathbf{B^\text{R}} \equiv \frac{\delta_t}{2} \sum_{k = q}^{p} (2 - \delta_{p,k} -  \delta_{k,q})\mathbf{A^\text{R}}_{p,k} \mathbf{B^\text{R}}_{k,q}.
	\end{equation}
	This can be rewritten as 
	\begin{equation}
		\small
		\frac{ \mathbf{A^\text{R}}  *\mathbf{B^\text{R}}}{\delta_t} = (\mathbf{w} \odot \mathbf{A^\text{R}}) \cdot (\mathbf{w} \odot \mathbf{B^\text{R}}) +  \frac{\text{Dg}\left( \mathbf{A^\text{R}} \right) \text{Dg}\left( \mathbf{B^\text{R}} \right) }{4},
	\end{equation}
	where $\odot$ is the block-wise matrix product defined as $\left [ A \odot B \right ]_{p,q} \equiv A_{p,q} B_{p,q}$ with $p$, $q$ block indices.  $\text{Dg}\left( \mathbf{M} \right)$ is the block-diagonal matrix defined by  
	$\text{Dg}\left( \mathbf{M}\right)_{p,q} \equiv  \mathbf{M}_{p,q} \delta_{p,q}$,
	and $\mathbf{w}$  is the block matrix with entries:  $ [(1 - \delta_{p,q} /2 )]_{p,q}$.
	 Similar expressions of the discretized kernel product can be derived for the other relevant products, \text{i.e.} $A^\text{A} B^\text{A}$, $A^\text{K} B^\text{K}$, $A^\text{K} B^\text{R}$ and $A^\text{A} B^\text{K}$.  We can now turn to the matrix formulation of the discretized Dyson equation
	\begin{equation}
		G^\text{R} = g^\text{R} +F^\text{R} G^\text{R},
	\end{equation}
	where $F^\text{R} = g^\text{R} \Sigma^\text{R}$. Using the above results, we recast this equation as: 
\begin{align}
	\mathbf{w} \odot \mathbf{G^\text{R}} &= \mathbf{S} + \mathbf{g^\text{R}} + \delta_t \cdot (\mathbf{w} \odot \mathbf{F^\text{R}}) \cdot (\mathbf{w} \odot \mathbf{G^\text{R}}) \label{eq:dyson:compact}, \\
	\qquad \text{with }
	\mathbf{S} &\equiv \delta_t \cdot \left(\frac{\text{Dg}\left( \mathbf{F^\text{R}} \right)}{4} - \frac{1}{2}\right) \cdot \text{Dg}\left( \mathbf{G^\text{R}} \right).
\end{align}
Due to the causal nature of this equation, the diagonal terms of $\mathbf{G^\text{R}}$ can be quickly deduced from the diagonal elements of $\mathbf{F^\text{R}}$ and $\mathbf{g^\text{R}}$; hence, the terms $\mathbf{S}$ are known before solving for $\mathbf{G^\text{R}}$. As a result, this equation can be efficiently solved using matrix algebra methods.
\begin{table}[]
	\begin{tabular}{@{}llll@{}}
		\toprule
		& Operation  & \centering HSS matrix   & Dense matrix\\ \midrule
		& Direct compression  & $\mathcal{O}(r N^2)$ &  \\
		& Stochastic compression & $\mathcal{O}(r^2 N)$    \\
		& $A B$ &  $\mathcal{O}(r^2 N)$  & $\mathcal{O}(N^3)$  \\
		& $A \setminus B$ &  $\mathcal{O}(r^2 N)$  &    $\mathcal{O}(N^3)$ \\ 
		& $A + B$ &  $\mathcal{O}(r^2 N)$   & $\mathcal{O}( N^2)$    \\  \bottomrule
	\end{tabular}
	\caption{Time complexity of several matrix operations. $A$ and $B$ are $N \times N$ matrices, $r$ is their maximum HSS-rank of the involved matrices. For a review of HSS algorithms, see \cite{Massei2020}. \label{tab:hss:complexity}}
\end{table}

Eq.\,(\ref{eq:dyson:compact}) enables the direct use of matrix compression techniques. The success of the \emph{equation-of-motion} techniques, as well as recent development such as \cite{Tuovinen}, suggests that the Green function can often be well-approximated as the solution of a differential equation. By discretizing differential operators into banded matrices using an appropriate basis, it becomes evident that the inverse of such operators possesses low-rank off-diagonal blocks \cite{Vandebril2005}.
This property has received empirical validation from \cite{Kaye2021}.

Such matrices can be efficiently represented using recursive factorization methods that separate short and long timescales by splitting the matrix into diagonal and off-diagonal blocks. The off-diagonal blocks, which capture long timescales, are compressed with low-rank factorization techniques. Among the various compression methods that apply such strategy, we choose the \emph{hierarchically semi separable (HSS) }  factorization \cite{Chandrasekaran2006, Martinsson2011, Xia2010, Massei2020} using the \emph{Julia} package \emph{HssMatrices.jl} \cite{Bonev2021HssMatrices}.
The complexity of HSS matrix operations depends on both the matrix size and their HSS-ranks $r$, which is the maximal rank of the off-diagonal blocks, see tab.\,\ref{tab:hss:complexity}. After compression, most of the matrix algebra operations can be performed in $\mathcal{O}(r^2 N)$ space and time, significantly reducing the computational time and memory requirements when $r \ll N$. 
However, the direct compression of a matrix $A$ into its factorization $A_\text{HSS}$ cost of $\mathcal{O}(r N^2)$ operations. 
This bottleneck can be bypassed using stochastic compression methods such as the one proposed by \cite{Martinsson2011}. Supposing that any matrix entry can be evaluated in $\mathcal{O}(1)$, and that the matrix product $Ax$ and $A^\dagger x$ for any vector $x$ can be performed in $T_\text{mul}$ operations, the compression complexity is reduced to $\mathcal{O}(r T_\text{mul} + r^2 N)$ while mostly preserving the accuracy. Hence, when $T_\text{mul} = \mathcal{O}(N)$, the compression reaches $\mathcal{O}(r^2 N)$ complexity in time. 
Lastly, the element-wise product $(\mathbf{w} \odot \mathbf{A^\text{R}})$ can be performed either using a general element-wise product algorithm for HSS-matrices \cite{Massei2020} or by exploiting the fact that most of the elements of $\mathbf{w}$ are equal to one. The complexity of the resolution of the Dyson equation, including the factorization stage, is thus $\mathcal{O}(rT_\text{mul} + r^2N)$.

Implementing fast multiplications $Ax$ and $A'x$ requires exploiting the structure of the uncompressed matrix. Discretized stationary kernels are represented as block-Toeplitz matrices. Hence, the \emph{Fast Fourier Transform} allows for multiplications in $\mathcal{O}(N\log N)$, resulting in a compression complexity of $\mathcal{O}(rN\log N)$. When the matrices are sparse, a complexity of $\mathcal{O}(r^2N)$ is readily achieved. For kernels that are solutions of a differential equation, the sparse representation of the differential operator can be compressed and solved, achieving optimal complexity.
Therefore, a quasi-linear complexity is reached as long  as $r$ grows slowly enough with the problem size. 


\section{benchmark}

\begin{figure}
	\includegraphics[width=8.6cm]{ 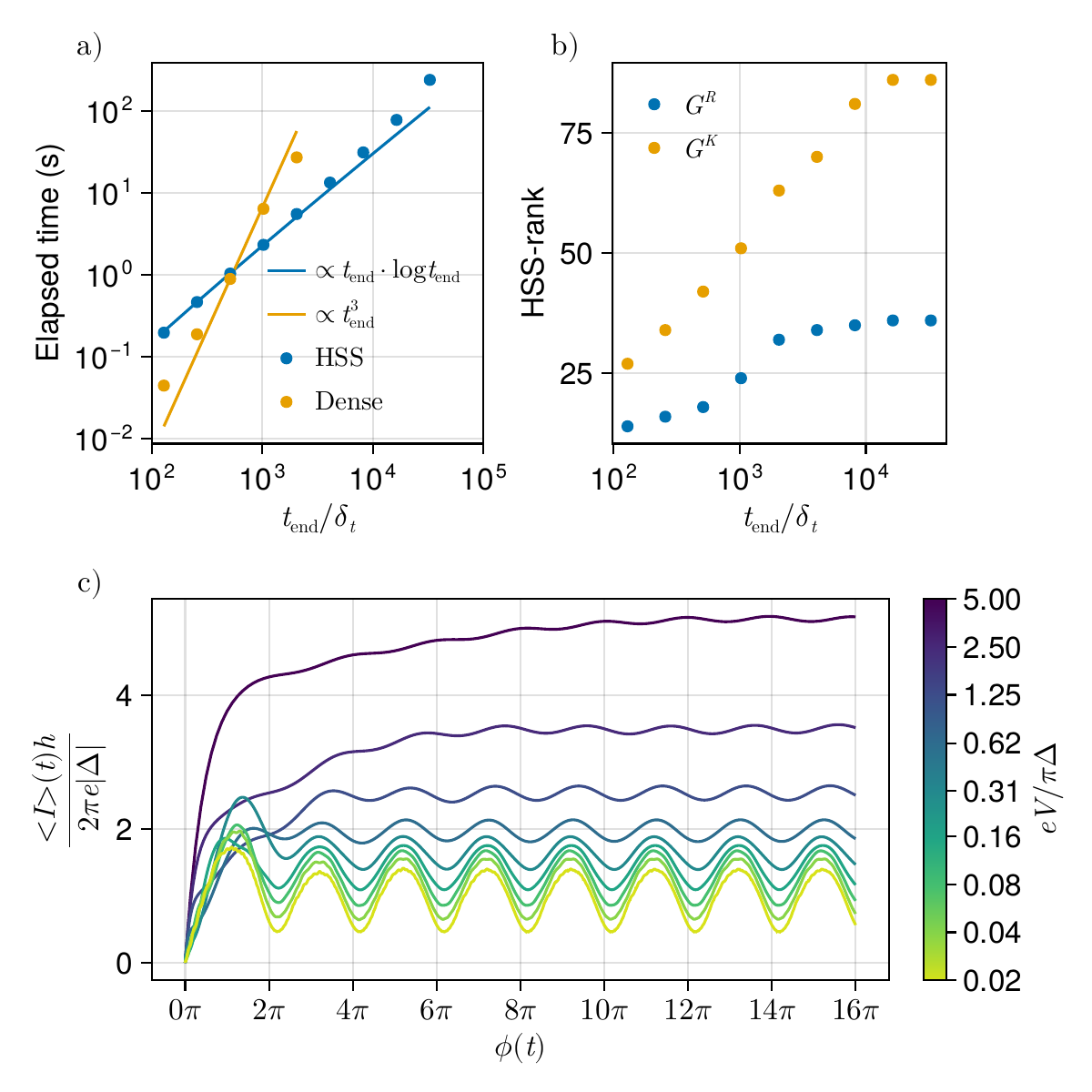}
	\caption{
		Solver benchmark for fixed time step $\delta_t$ with varying time horizon $t_\text{end}$ and voltage bias. The constant symmetric voltage  bias ensures that $\phi(t_{\text{end}}) = 16\pi$ with $\phi(0) = 0$.    The benchmark parameters are $\Gamma_L = \Gamma_R = 5 \Delta / \hbar$, $\delta_t = 0.025  h/\Delta, \beta = 10^2 \Delta^{-1}$. 
		\emph{a)} Solver runtime with and without compression.
		\emph{b)} HSS-ranks of the discretized full Green function.
		\emph{c)} Average current flowing from the dot to the right lead for various applied voltages.
		\label{fig:benchmark:fixed_dt} 
	}
\end{figure}

\begin{figure}
	\includegraphics[width=8.6cm]{ 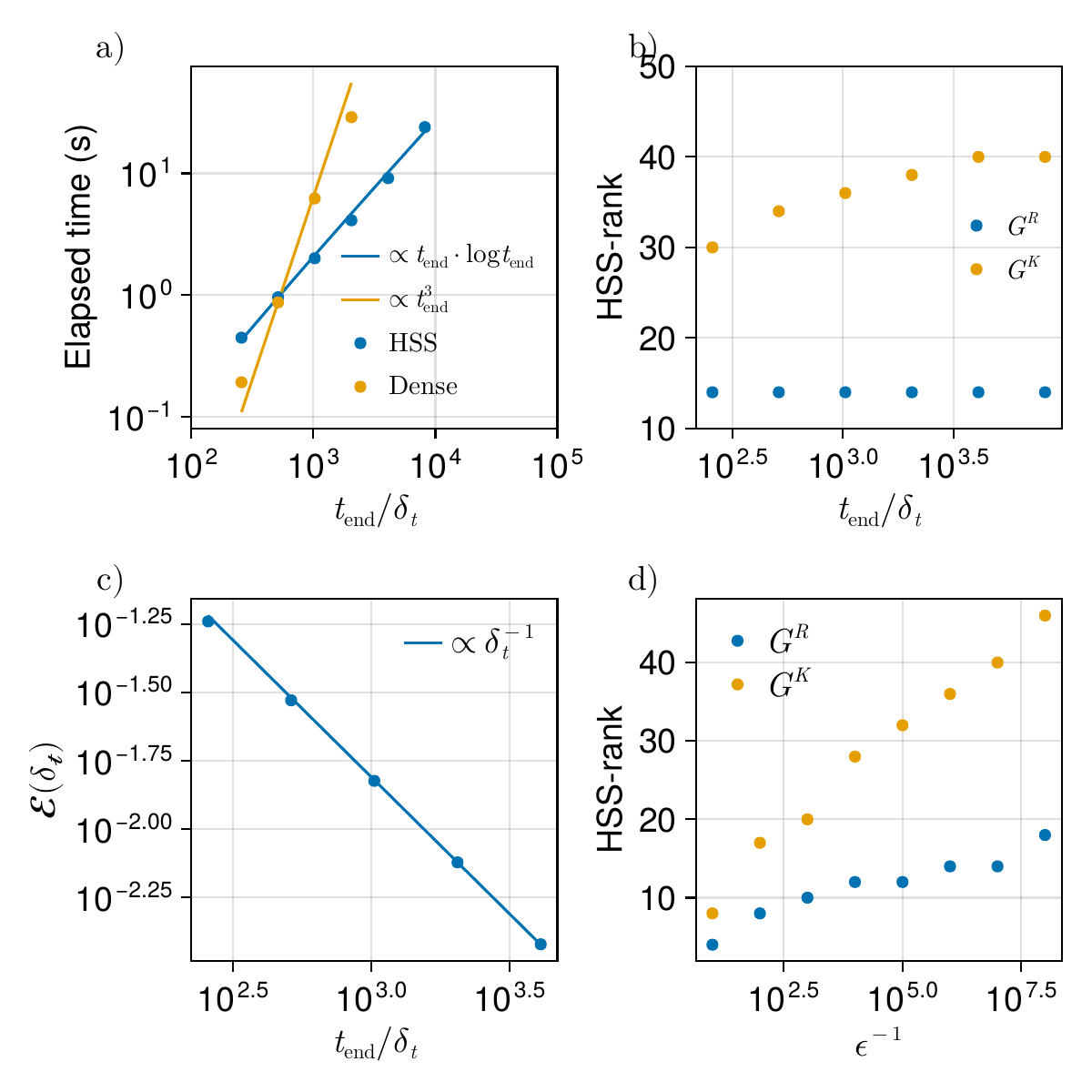}
	\caption{
		Solver benchmark for fixed time horizon $t_\text{end} = 3.2 h / \Delta$ with varying time step $\delta_t$.  The simulation parameters are: $\Gamma_L = \Gamma_R = 5 \Delta / \hbar$, $\beta = 10^2 \Delta^{-1}$. The constant voltage bias applied across the junction enforce that $\phi(t_{\text{end}}) = 16\pi$ with $\phi(0) = 0$. 
		\emph{a)} Solver runtime with and without compression.
		\emph{b)} HSS-ranks of the discretized full Green function.
		\emph{c) } Normalized error $\mathcal{E}$ in the current estimate as a function of $\delta_t$. 
		\emph{d)} Variation of the HSS-rank when varying the required accuracy $\epsilon = \epsilon_\text{rel} = \epsilon_{abs}$ of the HSS-representation for $\delta_t =  3.1\cdot 10^{-3} h/\Delta$.
		\label{fig:benchmark:convergence} 
	}
\end{figure}

We benchmarked this algorithm by simulating the out-of-equilibrium transport in a Josephson junction formed by a quantum dot (QD) connected to two superconducting leads as this system is well known \cite{Souto2020, Jacquet2020,Yeyati1997,Rogovin1974}. The solver implementation is hosted on github \cite{NonEquilibriumGreenFunction}. All the simulations were performed on a ultra 265K CPU with 48GB of RAM. The leads are superconductors described by the \emph{Barden-Cooper-Schrieffer} theory \cite{Tinkham2004} with a uniform real and positive order parameter $\Delta$. The operators $\psi_{\alpha,l,s}^\dagger$ create an electron of spin $s$ in mode $\alpha$ of lead $l$ with energy $ \varepsilon_{\alpha,l}$, while $d_s^\dagger$ creates an electron of spin $s$ on the dot with energy $\varepsilon_d$. $t_{\alpha,l}(t)$ represents the tunnel coupling between the dot and mode $\alpha$ of lead $l$. Using the usual Nambu spinor, where $d \equiv (d_\uparrow, d_\downarrow^\dagger)^T$ and $\Psi_{\alpha, l} \equiv ( \psi_{\alpha,l,\uparrow} , \psi_{\alpha,l,\downarrow}^\dagger )^T$, we can write the standard Hamiltonian as
$
 	H(t) = H_{L} + H_{D} + H_{T}(t)
$ \cite{Jacquet2020,Yeyati1997,Rogovin1974} , 
where
\begin{align}
	H_L
	&= \sum_{\alpha,l} \Psi_{\alpha,l}^\dagger \left(\sigma_x \Delta + \sigma_z \varepsilon_{\alpha,l}\right) \Psi_{\alpha,l},
	\\
	H_D &= d^\dagger \sigma_z \varepsilon_d d,
	\\
	H_T(t) &= \sum_{\alpha,l} \left\{ \Psi_{\alpha,l}^\dagger  \mathcal{T}_{\alpha,l}(t)  d + \text{H.c} \right\},
\end{align}
with $\mathcal{T}_{\alpha,l} \equiv t_{\alpha,l} \sigma_z e^{i\sigma_z \phi_{\alpha,l}(t)/2}$ being the time-dependent tunnel matrix, $\sigma_i$ the Pauli matrices acting in the Nambu space, the transmission phase $\phi_l(t) \equiv \frac{2e}{\hbar} \int^t{V_l(\tau)} d\tau $, $-e$ the charge of an electron and $V_l(\tau)$ the voltage bias from the QD to the lead $l$.  Employing the quantum field formalism \cite{Rammer2007, Altland2010}, we integrate out the infinite leads, resulting in an effective description of the dot in terms of its bare Green function $g_d$ dressed by the self-energies $\Sigma_l$ induced by the leads
\begin{equation}
	\Sigma_l = \hbar^{-1}\sum_{\alpha, \alpha'}\mathcal{T}_{\alpha,l}(t)^\dagger \cdot g_{(\alpha,l),(\alpha',l)}(t,t') \cdot \mathcal{T}_{\alpha',l}(t')^\dagger.
\end{equation}
In the flat-band limit, and upon absorbing the density of states into the tunneling coefficient, the  retarded Green functions of the leads are
\begin{equation}
	g_{\alpha}^{R}(t)
	= -i  \pi \delta(t) \sigma_0 + \pi  \Delta \left\{J_0( \frac{\Delta t}{\hbar} )\sigma_x + i J_1( \frac{\Delta t}{\hbar}) \sigma_0 \right\}, 
\end{equation} 
with $J_k$ the Bessel functions and $\sigma_0$ the identity matrix in the Nambu space. The dot's retarded Green function is
\begin{equation}
	g_{\text{dot}}^\text{R}(t) = -i e^{- i\sigma_z \varepsilon_d   t / \hbar }. 
\end{equation}
The average electric current flowing from the dot to the lead $l$ is \cite{Jacquet2020}
\begin{multline}
	\left \langle I_l(t) \right \rangle =  \frac{e }{ 2 \hbar} \text{tr}^{NK}\left\{ \tau_z \sigma_z
	\int \left[
	G(t,\tau) \Sigma_l(\tau,t) 	\right. \right.  \\ 
 \left. \left.	- \Sigma_l(t,\tau) G(\tau,t) \right] \text{d} \tau 
	\right\}
\end{multline}
where $G$ is the full Green function of the dot, $\text{tr}^\text{NK}$ is the trace over the Keldysh and Nambu indices, and $\tau_z$ is an operator acting on the Nambu space. Using the same Keldysh space basis as \cref{eq:basis:def}, its matrix representation is
\begin{equation}
	\tau_z =\frac{1}{2}\begin{pmatrix}
		1 & 2 \\
		2 & 3
	\end{pmatrix}.
\end{equation}

The evaluation of $\left \langle I_l(t) \right \rangle $ is performed using the fast kernel product described above; hence, it is deduced in $\mathcal{O}(r^2 N)$ from the full Green function. We suppose that the bare QD is half filled, \textit{i.e.} $g_{\text{dot}}^\text{K}(t) = 0$, and that the leads are initially at thermal equilibrium, hence the bare lead Keldysh Green functions satisfy
\begin{equation}
 g_{l}^\text{K} = g_{l}^\text{R} \cdot \rho_\beta - \rho_\beta  \cdot {g_{l}^\text{A}}, 
\end{equation}
	where $\rho_\beta(t,t')  \equiv -i \beta^{-1}  \text{csch} \left\{ \pi (t-t') \beta^{-1}\right\}$ and $\beta \equiv 1/k_B T$. We name the two superconductor leads left ($L$) and right ($R$).  For $t \geq 0$, the junction is drive out-of-equilibrium by symmetric voltage bias, \text{i.e.} $V_{L/R}(t) = \pm V/2 $. The phase difference $\phi(t) = \phi_R(t) - \phi_L(t)$ satisfies $\phi(0) = 0$.  We neglect any dependence of the tunnel coupling on the mode index $t_{\alpha,l} = t_{\alpha}$ and the tunnelling rate associate to the lead $l$ is $\Gamma_{l} =  \pi |t_{l}|^2/\hbar$. 
Such Josephson junctions host a pair of discrete particle-hole symmetric bound states, known as Andreev bound states (ABS), which are entirely detached from the continuum outside the superconducting gap and exhibit avoided crossing behaviour at the Fermi energy  when the dot is not resonant. Each of these states carries an opposite current \cite{Yeyati2003, Yeyati1997}.   We set $\Gamma_{\alpha} = 5\Delta / \hbar$, $\beta = 10^2 \Delta^{-1}$ and $ \epsilon_d = 0$.

For the first set of experiments, we impose $\phi(t_{\text{end}}) = 16\pi$  by setting $V = 2\hbar \phi(t_{\text{end}})/ e t_{\text{end}}$.  Thus, we have $\phi(t) = 16\pi t / t_{\text{end}}$. 
We measure the solver runtime for increased simulation length $t_\text{end}$, keeping $\delta_t$ constant as shown in  \cref{fig:benchmark:fixed_dt}. At $\phi \ll 1$ the states are symmetrically occupied, and the current vanishes. However, the voltage bias induces non-adiabatic transitions that couple the ABSs to each other and to the continuum outside the superconducting gap, driving the system into a non-stationary steady state \cite{Yeyati2003, Yeyati1997, BLamic2020}. In the steady state, the ABSs occupation is not symmetric, hence a net current flow across the junction. The HSS-rank of the retarded Green function saturates, while the one of the kinetic Green function grows slowly enough to ensure quasi-linear solving time in $t_{\text{end}} / \delta_t$.

 The simulation error arises from the interplay between discretization error and compression error. When the compression tolerance is sufficiently small, the error is dominated by the discretization error $\epsilon(\delta_t)$, which can be assumed to have the form
 	\begin{equation}
 		\epsilon(\delta_t) = \sum_{p \geq \alpha + 1} a_p \, \delta_t^p + o(\delta_t^n),
 	\end{equation}
 	where $\alpha \in \mathbb{N}^+$ is the convergence rate. By performing $n_\text{sim}$ simulations with varying time steps $\delta_{t, 0 < i \leq n_\text{sim}}$, one can evaluate $n_\text{sim} - 1$ terms of this expansion and  
construct a simulation result whose discretization error $\epsilon(\delta_{t,1}, ..., \delta_{t,n})$ is of the form 
 	  \begin{equation}
 	 	\epsilon(\delta_{t,1}, ..., \delta_{t,n}) = \sum_{p \geq \alpha + n_\text{sim}} a_p' \delta_t^p  + o(\delta_t^n),
 	 \end{equation}
 Hence, this increases the method convergence rate from $\alpha$ to at least $\alpha + n_\text{sim}$. This classical method is known as Richardson acceleration \cite{numericalRecipies}.
 To evaluate the solver convergence rate, we first compute a high-accuracy reference current estimate $\left\langle I_\text{ref} \right\rangle(t)$ using Richardson acceleration
 	\begin{equation}
 		\left\langle I_\text{ref} \right\rangle(t) \equiv \frac{\left\langle I_{\delta^*_t}\right\rangle(t) - 6 \left\langle I_{\delta^*_t /  2}\right\rangle(t)  + 8  \left\langle I_{\delta^*_t / 4}\right\rangle(t)} {3},
 	\end{equation}
 	where $\delta^*_t = 3.9\cdot 10^{-4}\Delta / \hbar$ and  $\left\langle I_{\delta_t}(t)\right\rangle$  the current evaluated for the time step $\delta_t$.  By construction, this estimate as a convergence rate of at least $3$ in $\delta^*_t$.
We then define the convergence error  for $\delta_t > \delta_t^*$ as
\begin{equation}
\mathcal{E}\left( \delta_t \right) \equiv \frac{\int|\left\langle I_{\delta_t}(t)\right\rangle-\left\langle I_\text{ref}(t)\right\rangle|^2\text{d}t}{\int|\left\langle I_\text{ref}(t)\right\rangle|^2\text{d}t}, 
\end{equation}
the evolution of this quantity with $\delta_t$ is shown in subplot \emph{c)} of \cref{fig:benchmark:convergence}, from which we extract a convergence rate $\alpha = 1$, \emph{i.e.} $\mathcal{E} \propto \delta_t^{-1}$. Although a first-order exact quadrature typically induces a second-order convergence rate, we attribute this slower convergence to the discontinuity of $\rho_\beta$. It can be overcome by using Richardson acceleration, or by improving the quadrature. We also observe when $\delta_t$ is reduced that the HSS-rank of $G^\text{R}$ remains constant, while the one of $G^\text{K}$ grows slowly enough to maintain quasi-linear complexity, see \cref{fig:benchmark:convergence}\emph{.a)} and \cref{fig:benchmark:convergence}\emph{.b)}. Using the same setup, we evaluate the algorithm dependency on the absolute tolerance $\epsilon_\text{abs}$ and the relative tolerance $\epsilon_\text{rel}$ use for the HSS compression. These parameters act as cutoffs in the selection of the singular values of the off diagonal blocks. Increasing the tolerance results in a decrease in the HSS ranks, thereby speeding up the solver. They should be chosen small enough to not penalize the solving accuracy while being large enough to not needlessly slow down the solver. In \cref{fig:benchmark:convergence}\emph{.d)}, we observe the dependence of HSS-rank with respect to $\epsilon_\text{abs}$ and $\epsilon_\text{rel}$. The exponential convergence of the error as the HSS-rank increases shows that compression can be used even when high accuracy is needed.

\begin{figure}
	\centering
	\includegraphics[width=8.6cm]{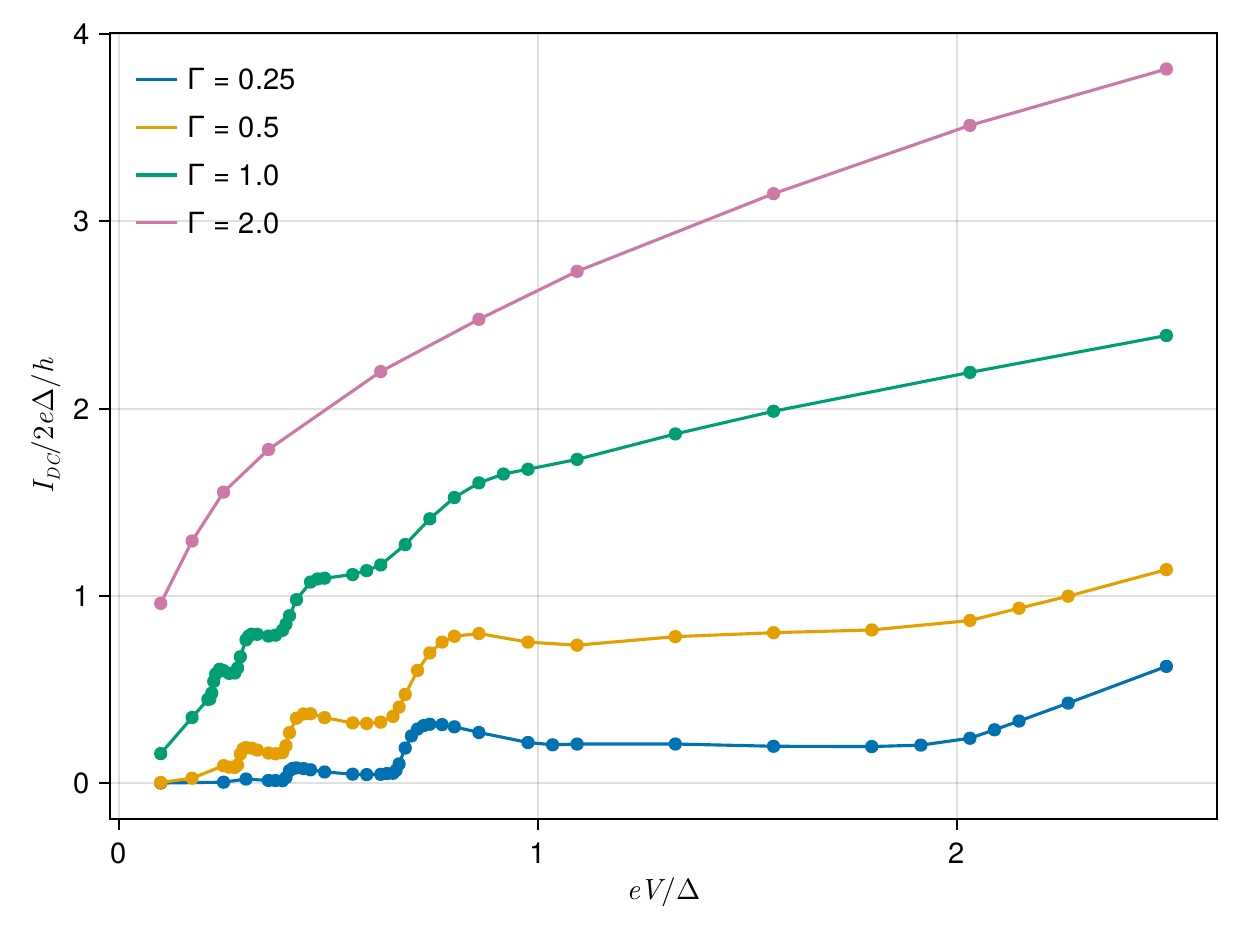}
	\caption{
		I-V characteristics of a voltage-biased quantum-dot Josephson junction with symmetric coupling $\Gamma = \Gamma_L = \Gamma_R$ at resonance $\varepsilon = 0$. Transient simulations run from $t = 0$ to $t_\text{end} = \max(200 h/\Delta, 20 T_\text{J})$ with compression tolerances $\epsilon_\text{rel} = \epsilon_\text{abs} = 10^{-6}$ in simulation units. Results reproduce the classical behavior from.~\cite{Yeyati1997}.
		\label{fig:steady_state} 
	}
\end{figure}

We extract the low-frequency averaged current, $I_\text{LF}(t)$, by averaging the instantaneous current $\langle I_l(t) \rangle$ over two Josephson periods, $2 T_\text{J} = 2\pi / (e V \hbar)$, accounting for the fractional Josephson radiation characteristic of a quantum dot Josephson junction at resonance \cite{BLamic2020}
\begin{equation}
	I^{\text{LF}}_{\delta_t}(t) \equiv \frac{1}{2 T_\text{J}} \int_{t - 2 T_\text{J}}^{t} \langle I_{\delta_t}(t') \rangle\, dt'.
\end{equation}
 The system is simulated from to $t_0 = 0$ to $t_\text{end} = \max(200\, \hbar/\Delta,\ 20\, T_\text{J})$
 to accommodate the different timescales across voltage regimes. At low voltage, the longest timescale is the Josephson period $T_\text{J}$, where interactions between the ABSs and the continuum occur primarily at anti-crossings, requiring simulation over many Josephson periods. At higher bias, however, the stronger couplings allow the system to reach steady state more rapidly, eliminating the need for such extended simulations. 
We use an adaptive multi-step Richardson extrapolation to both increase the result accuracy and provide an error estimate. The implementation is part of the solver module see \cite{NonEquilibriumGreenFunction}.  The step size is refined, until the absolute or relative error estimate are below $10^{-2}$ in simulation units. The initial time step is set to $\min(0.3 T_\text{J}, 0.5 /\Gamma )$ to ensure proper resolution of the fast timescale.
 
As the ABSs are separated from the continuum by a finite gap $\Delta_\text{ABS}$. For $V \ll f(\Delta_\text{ABS})$, the ABSs become effectively decoupled from the continuum \cite{BLamic2020, Yeyati2003}. In the absence of relaxation mechanisms, their long-time occupations depend on initial conditions, implying no well-defined steady state exists. Nevertheless, $I_\text{LF}(t)$ rapidly approaches a stationary regime, enabling extraction  of meaningful current-voltage characteristics even at low voltages. This behavior is illustrated in \cref{fig:steady_state}, which reproduces the classical results from \cite{Yeyati1997}.

	\section{Conclusion}
	
		By carefully crafting a discretization of the Dyson equation and by using modern matrix compression codes, we solved the out-of-equilibrium Dyson equation in quasi-linear time and space. Benchmarks demonstrate quasilinear scaling, with the measured runtime exhibiting $\mathcal{O}(N \log N)$ behavior when using \emph{HSS} compression, in contrast to the expected $\mathcal{O}(N^3)$ trend observed in reference simulations.  Utilizing a desktop, we accessed parameter regimes that are typically beyond the reach of a single-CPU machine. 
		This quasi-linear complexity critically depends on the effectiveness of HSS compression in representing the Green functions and self-energies, which we expect to generally hold, as discussed earlier. Consequently, given the existence of scalable and efficient implementations of HSS compression or similar methods, and considering their robustness, we anticipate that this solving technique will be capable of addressing extremely large problems, provided that the compressed representations of the kernels fit in memory.
		
		This strategy might be extended to solve quantum field equations involving higher-order correlation functions, whose resolution has so far been impeded by the absence of efficient methods. Such an extension would pave a new way for the simulation of interacting many-body systems far from equilibrium in the transient regimes.
		
	\section*{Acknowledgments}
	
	The author would like to thank Manuel Houzet and Julia S. Meyer for their valuable feedback, and Cécile X. Yu for her proofreading and support.
	
	\section*{Data availability}
	The datasets and scripts used to produce all figures and simulations in this work are publicly accessible at \cite{FastDysonSolver}.
	
	\bibliography{biblio}

\begin{thebibliography}{32}%
\makeatletter
\providecommand \@ifxundefined [1]{%
 \@ifx{#1\undefined}
}%
\providecommand \@ifnum [1]{%
 \ifnum #1\expandafter \@firstoftwo
 \else \expandafter \@secondoftwo
 \fi
}%
\providecommand \@ifx [1]{%
 \ifx #1\expandafter \@firstoftwo
 \else \expandafter \@secondoftwo
 \fi
}%
\providecommand \natexlab [1]{#1}%
\providecommand \enquote  [1]{``#1''}%
\providecommand \bibnamefont  [1]{#1}%
\providecommand \bibfnamefont [1]{#1}%
\providecommand \citenamefont [1]{#1}%
\providecommand \href@noop [0]{\@secondoftwo}%
\providecommand \href [0]{\begingroup \@sanitize@url \@href}%
\providecommand \@href[1]{\@@startlink{#1}\@@href}%
\providecommand \@@href[1]{\endgroup#1\@@endlink}%
\providecommand \@sanitize@url [0]{\catcode `\\12\catcode `\$12\catcode
  `\&12\catcode `\#12\catcode `\^12\catcode `\_12\catcode `\%12\relax}%
\providecommand \@@startlink[1]{}%
\providecommand \@@endlink[0]{}%
\providecommand \url  [0]{\begingroup\@sanitize@url \@url }%
\providecommand \@url [1]{\endgroup\@href {#1}{\urlprefix }}%
\providecommand \urlprefix  [0]{URL }%
\providecommand \Eprint [0]{\href }%
\providecommand \doibase [0]{https://doi.org/}%
\providecommand \selectlanguage [0]{\@gobble}%
\providecommand \bibinfo  [0]{\@secondoftwo}%
\providecommand \bibfield  [0]{\@secondoftwo}%
\providecommand \translation [1]{[#1]}%
\providecommand \BibitemOpen [0]{}%
\providecommand \bibitemStop [0]{}%
\providecommand \bibitemNoStop [0]{.\EOS\space}%
\providecommand \EOS [0]{\spacefactor3000\relax}%
\providecommand \BibitemShut  [1]{\csname bibitem#1\endcsname}%
\let\auto@bib@innerbib\@empty
\bibitem [{\citenamefont {Altland}\ and\ \citenamefont
  {Simons}(2010)}]{Altland2010}%
  \BibitemOpen
  \bibfield  {author} {\bibinfo {author} {\bibfnamefont {A.}~\bibnamefont
  {Altland}}\ and\ \bibinfo {author} {\bibfnamefont {B.~D.}\ \bibnamefont
  {Simons}},\ }\href {https://doi.org/10.1017/CBO9780511789984} {\emph
  {\bibinfo {title} {Condensed Matter Theories}}},\ Vol.~\bibinfo {volume}
  {23}\ (\bibinfo  {publisher} {Cambridge University Press},\ \bibinfo
  {address} {Cambridge},\ \bibinfo {year} {2010})\ pp.\ \bibinfo {pages}
  {1--478}\BibitemShut {NoStop}%
\bibitem [{\citenamefont {Rammer}(2007)}]{Rammer2007}%
  \BibitemOpen
  \bibfield  {author} {\bibinfo {author} {\bibfnamefont {J.}~\bibnamefont
  {Rammer}},\ }\href {https://doi.org/10.1017/CBO9780511618956} {\emph
  {\bibinfo {title} {Quantum Field Theory of Non-Equilibrium States}}},\ Vol.\
  \bibinfo {volume} {9780521874}\ (\bibinfo  {publisher} {Cambridge University
  Press},\ \bibinfo {address} {Cambridge},\ \bibinfo {year} {2007})\ pp.\
  \bibinfo {pages} {1--536}\BibitemShut {NoStop}%
\bibitem [{\citenamefont {Profumo}\ \emph {et~al.}(2015)\citenamefont
  {Profumo}, \citenamefont {Groth}, \citenamefont {Messio}, \citenamefont
  {Parcollet},\ and\ \citenamefont {Waintal}}]{Profumo2015}%
  \BibitemOpen
  \bibfield  {author} {\bibinfo {author} {\bibfnamefont {R.~E.~V.}\
  \bibnamefont {Profumo}}, \bibinfo {author} {\bibfnamefont {C.}~\bibnamefont
  {Groth}}, \bibinfo {author} {\bibfnamefont {L.}~\bibnamefont {Messio}},
  \bibinfo {author} {\bibfnamefont {O.}~\bibnamefont {Parcollet}},\ and\
  \bibinfo {author} {\bibfnamefont {X.}~\bibnamefont {Waintal}},\ }\bibfield
  {title} {\bibinfo {title} {Quantum monte carlo for correlated
  out-of-equilibrium nanoelectronic devices},\ }\href
  {https://doi.org/10.1103/PhysRevB.91.245154} {\bibfield  {journal} {\bibinfo
  {journal} {Physical Review B}\ }\textbf {\bibinfo {volume} {91}},\ \bibinfo
  {pages} {245154} (\bibinfo {year} {2015})}\BibitemShut {NoStop}%
\bibitem [{\citenamefont {Kozik}\ \emph {et~al.}(2010)\citenamefont {Kozik},
  \citenamefont {Houcke}, \citenamefont {Gull}, \citenamefont {Pollet},
  \citenamefont {Prokof'ev}, \citenamefont {Svistunov},\ and\ \citenamefont
  {Troyer}}]{Kozik2010}%
  \BibitemOpen
  \bibfield  {author} {\bibinfo {author} {\bibfnamefont {E.}~\bibnamefont
  {Kozik}}, \bibinfo {author} {\bibfnamefont {K.~V.}\ \bibnamefont {Houcke}},
  \bibinfo {author} {\bibfnamefont {E.}~\bibnamefont {Gull}}, \bibinfo {author}
  {\bibfnamefont {L.}~\bibnamefont {Pollet}}, \bibinfo {author} {\bibfnamefont
  {N.}~\bibnamefont {Prokof'ev}}, \bibinfo {author} {\bibfnamefont
  {B.}~\bibnamefont {Svistunov}},\ and\ \bibinfo {author} {\bibfnamefont
  {M.}~\bibnamefont {Troyer}},\ }\bibfield  {title} {\bibinfo {title}
  {Diagrammatic monte carlo for correlated fermions},\ }\href
  {https://doi.org/10.1209/0295-5075/90/10004} {\bibfield  {journal} {\bibinfo
  {journal} {EPL (Europhysics Letters)}\ }\textbf {\bibinfo {volume} {90}},\
  \bibinfo {pages} {10004} (\bibinfo {year} {2010})}\BibitemShut {NoStop}%
\bibitem [{\citenamefont {Metzner}\ \emph {et~al.}(2012)\citenamefont
  {Metzner}, \citenamefont {Salmhofer}, \citenamefont {Honerkamp},
  \citenamefont {Meden},\ and\ \citenamefont {Schönhammer}}]{Metzner2012}%
  \BibitemOpen
  \bibfield  {author} {\bibinfo {author} {\bibfnamefont {W.}~\bibnamefont
  {Metzner}}, \bibinfo {author} {\bibfnamefont {M.}~\bibnamefont {Salmhofer}},
  \bibinfo {author} {\bibfnamefont {C.}~\bibnamefont {Honerkamp}}, \bibinfo
  {author} {\bibfnamefont {V.}~\bibnamefont {Meden}},\ and\ \bibinfo {author}
  {\bibfnamefont {K.}~\bibnamefont {Schönhammer}},\ }\bibfield  {title}
  {\bibinfo {title} {Functional renormalization group approach to correlated
  fermion systems},\ }\href {https://doi.org/10.1103/RevModPhys.84.299}
  {\bibfield  {journal} {\bibinfo  {journal} {Reviews of Modern Physics}\
  }\textbf {\bibinfo {volume} {84}},\ \bibinfo {pages} {299} (\bibinfo {year}
  {2012})}\BibitemShut {NoStop}%
\bibitem [{\citenamefont {Dong}\ \emph {et~al.}(2020)\citenamefont {Dong},
  \citenamefont {Zgid}, \citenamefont {Gull},\ and\ \citenamefont
  {Strand}}]{Dong2020}%
  \BibitemOpen
  \bibfield  {author} {\bibinfo {author} {\bibfnamefont {X.}~\bibnamefont
  {Dong}}, \bibinfo {author} {\bibfnamefont {D.}~\bibnamefont {Zgid}}, \bibinfo
  {author} {\bibfnamefont {E.}~\bibnamefont {Gull}},\ and\ \bibinfo {author}
  {\bibfnamefont {H.~U.}\ \bibnamefont {Strand}},\ }\bibfield  {title}
  {\bibinfo {title} {{Legendre-spectral Dyson equation solver with
  super-exponential convergence}},\ }\bibfield  {journal} {\bibinfo  {journal}
  {Journal of Chemical Physics}\ }\textbf {\bibinfo {volume} {152}},\ \href
  {https://doi.org/10.1063/5.0003145} {10.1063/5.0003145} (\bibinfo {year}
  {2020}),\ \Eprint {https://arxiv.org/abs/2001.11603} {arXiv:2001.11603}
  \BibitemShut {NoStop}%
\bibitem [{\citenamefont {Talarico}\ \emph {et~al.}(2019)\citenamefont
  {Talarico}, \citenamefont {Maniscalco},\ and\ \citenamefont {{Lo
  Gullo}}}]{Talarico2019}%
  \BibitemOpen
  \bibfield  {author} {\bibinfo {author} {\bibfnamefont {N.~W.}\ \bibnamefont
  {Talarico}}, \bibinfo {author} {\bibfnamefont {S.}~\bibnamefont
  {Maniscalco}},\ and\ \bibinfo {author} {\bibfnamefont {N.}~\bibnamefont {{Lo
  Gullo}}},\ }\bibfield  {title} {\bibinfo {title} {{A Scalable Numerical
  Approach to the Solution of the Dyson Equation for the Non-Equilibrium
  Single-Particle Green's Function}},\ }\bibfield  {journal} {\bibinfo
  {journal} {Physica Status Solidi (B) Basic Research}\ }\textbf {\bibinfo
  {volume} {256}},\ \href {https://doi.org/10.1002/pssb.201800501}
  {10.1002/pssb.201800501} (\bibinfo {year} {2019}),\ \Eprint
  {https://arxiv.org/abs/1809.10111} {arXiv:1809.10111} \BibitemShut {NoStop}%
\bibitem [{\citenamefont {Kaye}\ and\ \citenamefont {Golez}(2021)}]{Kaye2021}%
  \BibitemOpen
  \bibfield  {author} {\bibinfo {author} {\bibfnamefont {J.}~\bibnamefont
  {Kaye}}\ and\ \bibinfo {author} {\bibfnamefont {D.}~\bibnamefont {Golez}},\
  }\bibfield  {title} {\bibinfo {title} {Low rank compression in the numerical
  solution of the nonequilibrium dyson equation},\ }\href
  {https://doi.org/10.21468/SciPostPhys.10.4.091} {\bibfield  {journal}
  {\bibinfo  {journal} {SciPost Physics}\ }\textbf {\bibinfo {volume} {10}},\
  \bibinfo {pages} {091} (\bibinfo {year} {2021})}\BibitemShut {NoStop}%
\bibitem [{\citenamefont {Kaye}\ and\ \citenamefont
  {Strand}(2021)}]{KayeEquilibrium}%
  \BibitemOpen
  \bibfield  {author} {\bibinfo {author} {\bibfnamefont {J.}~\bibnamefont
  {Kaye}}\ and\ \bibinfo {author} {\bibfnamefont {H.~U.~R.}\ \bibnamefont
  {Strand}},\ }\bibfield  {title} {\bibinfo {title} {A fast time domain solver
  for the equilibrium dyson equation},\ }\href@noop {} {\ \textbf {\bibinfo
  {volume} {2}},\ \bibinfo {pages} {1} (\bibinfo {year} {2021})}\BibitemShut
  {NoStop}%
\bibitem [{\citenamefont {Sch{\"{u}}ler}\ \emph {et~al.}(2020)\citenamefont
  {Sch{\"{u}}ler}, \citenamefont {Gole{\v{z}}}, \citenamefont {Murakami},
  \citenamefont {Bittner}, \citenamefont {Herrmann}, \citenamefont {Strand},
  \citenamefont {Werner},\ and\ \citenamefont {Eckstein}}]{NESSI}%
  \BibitemOpen
  \bibfield  {author} {\bibinfo {author} {\bibfnamefont {M.}~\bibnamefont
  {Sch{\"{u}}ler}}, \bibinfo {author} {\bibfnamefont {D.}~\bibnamefont
  {Gole{\v{z}}}}, \bibinfo {author} {\bibfnamefont {Y.}~\bibnamefont
  {Murakami}}, \bibinfo {author} {\bibfnamefont {N.}~\bibnamefont {Bittner}},
  \bibinfo {author} {\bibfnamefont {A.}~\bibnamefont {Herrmann}}, \bibinfo
  {author} {\bibfnamefont {H.~U.}\ \bibnamefont {Strand}}, \bibinfo {author}
  {\bibfnamefont {P.}~\bibnamefont {Werner}},\ and\ \bibinfo {author}
  {\bibfnamefont {M.}~\bibnamefont {Eckstein}},\ }\bibfield  {title} {\bibinfo
  {title} {{NESSi: The Non-Equilibrium Systems Simulation package}},\ }\href
  {https://doi.org/10.1016/j.cpc.2020.107484} {\bibfield  {journal} {\bibinfo
  {journal} {Computer Physics Communications}\ }\textbf {\bibinfo {volume}
  {257}},\ \bibinfo {pages} {107484} (\bibinfo {year} {2020})}\BibitemShut
  {NoStop}%
\bibitem [{\citenamefont {Kloss}\ \emph {et~al.}(2021)\citenamefont {Kloss},
  \citenamefont {Weston}, \citenamefont {Gaury}, \citenamefont {Rossignol},
  \citenamefont {Groth},\ and\ \citenamefont {Waintal}}]{kloss2021tkwant}%
  \BibitemOpen
  \bibfield  {author} {\bibinfo {author} {\bibfnamefont {T.}~\bibnamefont
  {Kloss}}, \bibinfo {author} {\bibfnamefont {J.}~\bibnamefont {Weston}},
  \bibinfo {author} {\bibfnamefont {B.}~\bibnamefont {Gaury}}, \bibinfo
  {author} {\bibfnamefont {B.}~\bibnamefont {Rossignol}}, \bibinfo {author}
  {\bibfnamefont {C.}~\bibnamefont {Groth}},\ and\ \bibinfo {author}
  {\bibfnamefont {X.}~\bibnamefont {Waintal}},\ }\bibfield  {title} {\bibinfo
  {title} {Tkwant: a software package for time-dependent quantum transport},\
  }\href@noop {} {\bibfield  {journal} {\bibinfo  {journal} {New Journal of
  Physics}\ }\textbf {\bibinfo {volume} {23}},\ \bibinfo {pages} {023025}
  (\bibinfo {year} {2021})}\BibitemShut {NoStop}%
\bibitem [{\citenamefont {Tuovinen}\ \emph {et~al.}(2023)\citenamefont
  {Tuovinen}, \citenamefont {Pavlyukh}, \citenamefont {Perfetto},\ and\
  \citenamefont {Stefanucci}}]{Tuovinen}%
  \BibitemOpen
  \bibfield  {author} {\bibinfo {author} {\bibfnamefont {R.}~\bibnamefont
  {Tuovinen}}, \bibinfo {author} {\bibfnamefont {Y.}~\bibnamefont {Pavlyukh}},
  \bibinfo {author} {\bibfnamefont {E.}~\bibnamefont {Perfetto}},\ and\
  \bibinfo {author} {\bibfnamefont {G.}~\bibnamefont {Stefanucci}},\ }\bibfield
   {title} {\bibinfo {title} {Time-linear quantum transport simulations with
  correlated nonequilibrium green's functions},\ }\href
  {https://doi.org/10.1103/PhysRevLett.130.246301} {\bibfield  {journal}
  {\bibinfo  {journal} {Phys. Rev. Lett.}\ }\textbf {\bibinfo {volume} {130}},\
  \bibinfo {pages} {246301} (\bibinfo {year} {2023})}\BibitemShut {NoStop}%
\bibitem [{\citenamefont {Ortega-Taberner}\ \emph {et~al.}(2023)\citenamefont
  {Ortega-Taberner}, \citenamefont {Jauho},\ and\ \citenamefont
  {Paaske}}]{Ortega-Taberner2023}%
  \BibitemOpen
  \bibfield  {author} {\bibinfo {author} {\bibfnamefont {C.}~\bibnamefont
  {Ortega-Taberner}}, \bibinfo {author} {\bibfnamefont {A.-P.}\ \bibnamefont
  {Jauho}},\ and\ \bibinfo {author} {\bibfnamefont {J.}~\bibnamefont
  {Paaske}},\ }\bibfield  {title} {\bibinfo {title} {Anomalous josephson
  current through a driven double quantum dot},\ }\href
  {https://doi.org/10.1103/PhysRevB.107.115165} {\bibfield  {journal} {\bibinfo
   {journal} {Physical Review B}\ }\textbf {\bibinfo {volume} {107}},\ \bibinfo
  {pages} {115165} (\bibinfo {year} {2023})}\BibitemShut {NoStop}%
\bibitem [{\citenamefont {Venitucci}\ \emph {et~al.}(2018)\citenamefont
  {Venitucci}, \citenamefont {Feinberg}, \citenamefont {Mélin},\ and\
  \citenamefont {Douçot}}]{Venitucci2018}%
  \BibitemOpen
  \bibfield  {author} {\bibinfo {author} {\bibfnamefont {B.}~\bibnamefont
  {Venitucci}}, \bibinfo {author} {\bibfnamefont {D.}~\bibnamefont {Feinberg}},
  \bibinfo {author} {\bibfnamefont {R.}~\bibnamefont {Mélin}},\ and\ \bibinfo
  {author} {\bibfnamefont {B.}~\bibnamefont {Douçot}},\ }\bibfield  {title}
  {\bibinfo {title} {Nonadiabatic josephson current pumping by chiral microwave
  irradiation},\ }\href {https://doi.org/10.1103/PhysRevB.97.195423} {\bibfield
   {journal} {\bibinfo  {journal} {Physical Review B}\ }\textbf {\bibinfo
  {volume} {97}},\ \bibinfo {pages} {195423} (\bibinfo {year}
  {2018})}\BibitemShut {NoStop}%
\bibitem [{\citenamefont {Fornberg}\ and\ \citenamefont
  {Reeger}(2019)}]{Fornberg2019}%
  \BibitemOpen
  \bibfield  {author} {\bibinfo {author} {\bibfnamefont {B.}~\bibnamefont
  {Fornberg}}\ and\ \bibinfo {author} {\bibfnamefont {J.~A.}\ \bibnamefont
  {Reeger}},\ }\bibfield  {title} {\bibinfo {title} {An improved gregory-like
  method for 1-d quadrature},\ }\href
  {https://doi.org/10.1007/s00211-018-0992-0} {\bibfield  {journal} {\bibinfo
  {journal} {Numerische Mathematik}\ }\textbf {\bibinfo {volume} {141}},\
  \bibinfo {pages} {1} (\bibinfo {year} {2019})}\BibitemShut {NoStop}%
\bibitem [{\citenamefont {Lamic}(2025{\natexlab{a}})}]{Supplement}%
  \BibitemOpen
  \bibfield  {author} {\bibinfo {author} {\bibfnamefont {B.}~\bibnamefont
  {Lamic}},\ }\href@noop {} {\bibinfo {title} {Supplemental material for:
  Solving the transient dyson equation with quasilinear complexity via matrix
  compression}},\ \bibinfo {howpublished} {Available as supplemental
  information with the article} (\bibinfo {year}
  {2025}{\natexlab{a}})\BibitemShut {NoStop}%
\bibitem [{\citenamefont {Massei}\ \emph {et~al.}(2020)\citenamefont {Massei},
  \citenamefont {Robol},\ and\ \citenamefont {Kressner}}]{Massei2020}%
  \BibitemOpen
  \bibfield  {author} {\bibinfo {author} {\bibfnamefont {S.}~\bibnamefont
  {Massei}}, \bibinfo {author} {\bibfnamefont {L.}~\bibnamefont {Robol}},\ and\
  \bibinfo {author} {\bibfnamefont {D.}~\bibnamefont {Kressner}},\ }\bibfield
  {title} {\bibinfo {title} {{Hm-toolbox: Matlab software for hodlr and HSS
  matrices}},\ }\href {https://doi.org/10.1137/19M1288048} {\bibfield
  {journal} {\bibinfo  {journal} {SIAM Journal on Scientific Computing}\
  }\textbf {\bibinfo {volume} {42}},\ \bibinfo {pages} {C43} (\bibinfo {year}
  {2020})},\ \Eprint {https://arxiv.org/abs/1909.07909} {arXiv:1909.07909}
  \BibitemShut {NoStop}%
\bibitem [{\citenamefont {Vandebril}\ \emph {et~al.}(2005)\citenamefont
  {Vandebril}, \citenamefont {Barel}, \citenamefont {Golub},\ and\
  \citenamefont {Mastronardi}}]{Vandebril2005}%
  \BibitemOpen
  \bibfield  {author} {\bibinfo {author} {\bibfnamefont {R.}~\bibnamefont
  {Vandebril}}, \bibinfo {author} {\bibfnamefont {M.~V.}\ \bibnamefont
  {Barel}}, \bibinfo {author} {\bibfnamefont {G.}~\bibnamefont {Golub}},\ and\
  \bibinfo {author} {\bibfnamefont {N.}~\bibnamefont {Mastronardi}},\
  }\bibfield  {title} {\bibinfo {title} {{A bibliography on semiseparable
  matrices}},\ }\href {https://doi.org/10.1007/s10092-005-0107-z} {\bibfield
  {journal} {\bibinfo  {journal} {Calcolo}\ }\textbf {\bibinfo {volume} {42}},\
  \bibinfo {pages} {249} (\bibinfo {year} {2005})}\BibitemShut {NoStop}%
\bibitem [{\citenamefont {Chandrasekaran}\ \emph {et~al.}(2006)\citenamefont
  {Chandrasekaran}, \citenamefont {Gu},\ and\ \citenamefont
  {Pals}}]{Chandrasekaran2006}%
  \BibitemOpen
  \bibfield  {author} {\bibinfo {author} {\bibfnamefont {S.}~\bibnamefont
  {Chandrasekaran}}, \bibinfo {author} {\bibfnamefont {M.}~\bibnamefont {Gu}},\
  and\ \bibinfo {author} {\bibfnamefont {T.}~\bibnamefont {Pals}},\ }\bibfield
  {title} {\bibinfo {title} {{A Fast $ULV$ Decomposition Solver for
  Hierarchically Semiseparable Representations}},\ }\href
  {https://doi.org/10.1137/S0895479803436652} {\bibfield  {journal} {\bibinfo
  {journal} {SIAM Journal on Matrix Analysis and Applications}\ }\textbf
  {\bibinfo {volume} {28}},\ \bibinfo {pages} {603} (\bibinfo {year}
  {2006})}\BibitemShut {NoStop}%
\bibitem [{\citenamefont {Martinsson}(2011)}]{Martinsson2011}%
  \BibitemOpen
  \bibfield  {author} {\bibinfo {author} {\bibfnamefont {P.~G.}\ \bibnamefont
  {Martinsson}},\ }\bibfield  {title} {\bibinfo {title} {A fast randomized
  algorithm for computing a hierarchically semiseparable representation of a
  matrix},\ }\href {https://doi.org/10.1137/100786617} {\bibfield  {journal}
  {\bibinfo  {journal} {SIAM Journal on Matrix Analysis and Applications}\
  }\textbf {\bibinfo {volume} {32}},\ \bibinfo {pages} {1251} (\bibinfo {year}
  {2011})}\BibitemShut {NoStop}%
\bibitem [{\citenamefont {Xia}\ \emph {et~al.}(2010)\citenamefont {Xia},
  \citenamefont {Chandrasekaran}, \citenamefont {Gu},\ and\ \citenamefont
  {Li}}]{Xia2010}%
  \BibitemOpen
  \bibfield  {author} {\bibinfo {author} {\bibfnamefont {J.}~\bibnamefont
  {Xia}}, \bibinfo {author} {\bibfnamefont {S.}~\bibnamefont {Chandrasekaran}},
  \bibinfo {author} {\bibfnamefont {M.}~\bibnamefont {Gu}},\ and\ \bibinfo
  {author} {\bibfnamefont {X.~S.}\ \bibnamefont {Li}},\ }\bibfield  {title}
  {\bibinfo {title} {{Fast algorithms for hierarchically semiseparable
  matrices}},\ }\href {https://doi.org/10.1002/nla.691} {\bibfield  {journal}
  {\bibinfo  {journal} {Numerical Linear Algebra with Applications}\ }\textbf
  {\bibinfo {volume} {17}},\ \bibinfo {pages} {953} (\bibinfo {year}
  {2010})}\BibitemShut {NoStop}%
\bibitem [{\citenamefont {Bonev}(2021)}]{Bonev2021HssMatrices}%
  \BibitemOpen
  \bibfield  {author} {\bibinfo {author} {\bibfnamefont {B.}~\bibnamefont
  {Bonev}},\ }\href {https://doi.org/10.5281/zenodo.4696465} {\bibinfo {title}
  {\text{HssMatrices.jl}: A julia package for hierarchically semi-separable
  matrices}} (\bibinfo {year} {2021}),\ \bibinfo {note}
  {\url{https://doi.org/10.5281/zenodo.4696465}}\BibitemShut {NoStop}%
\bibitem [{\citenamefont {Souto}(2020)}]{Souto2020}%
  \BibitemOpen
  \bibfield  {author} {\bibinfo {author} {\bibfnamefont {R.~S.}\ \bibnamefont
  {Souto}},\ }\href {https://doi.org/10.1007/978-3-030-36595-0} {\emph
  {\bibinfo {title} {Quench Dynamics in Interacting and Superconducting
  Nanojunctions}}}\ (\bibinfo  {publisher} {Springer International
  Publishing},\ \bibinfo {year} {2020})\BibitemShut {NoStop}%
\bibitem [{\citenamefont {Jacquet}\ \emph {et~al.}(2020)\citenamefont
  {Jacquet}, \citenamefont {Popoff}, \citenamefont {Imura}, \citenamefont
  {Rech}, \citenamefont {Jonckheere}, \citenamefont {Raymond}, \citenamefont
  {Zazunov},\ and\ \citenamefont {Martin}}]{Jacquet2020}%
  \BibitemOpen
  \bibfield  {author} {\bibinfo {author} {\bibfnamefont {R.}~\bibnamefont
  {Jacquet}}, \bibinfo {author} {\bibfnamefont {A.}~\bibnamefont {Popoff}},
  \bibinfo {author} {\bibfnamefont {K.~I.}\ \bibnamefont {Imura}}, \bibinfo
  {author} {\bibfnamefont {J.}~\bibnamefont {Rech}}, \bibinfo {author}
  {\bibfnamefont {T.}~\bibnamefont {Jonckheere}}, \bibinfo {author}
  {\bibfnamefont {L.}~\bibnamefont {Raymond}}, \bibinfo {author} {\bibfnamefont
  {A.}~\bibnamefont {Zazunov}},\ and\ \bibinfo {author} {\bibfnamefont
  {T.}~\bibnamefont {Martin}},\ }\bibfield  {title} {\bibinfo {title} {{Theory
  of nonequilibrium noise in general multiterminal superconducting hybrid
  devices: Application to multiple Cooper pair resonances}},\ }\bibfield
  {journal} {\bibinfo  {journal} {Physical Review B}\ }\textbf {\bibinfo
  {volume} {102}},\ \href {https://doi.org/10.1103/PhysRevB.102.064510}
  {10.1103/PhysRevB.102.064510} (\bibinfo {year} {2020})\BibitemShut {NoStop}%
\bibitem [{\citenamefont {Yeyati}\ \emph {et~al.}(1997)\citenamefont {Yeyati},
  \citenamefont {Cuevas}, \citenamefont {L{\'{o}}pez-D{\'{a}}valos},\ and\
  \citenamefont {Mart{\'{i}}n-Rodero}}]{Yeyati1997}%
  \BibitemOpen
  \bibfield  {author} {\bibinfo {author} {\bibfnamefont {A.~L.}\ \bibnamefont
  {Yeyati}}, \bibinfo {author} {\bibfnamefont {J.}~\bibnamefont {Cuevas}},
  \bibinfo {author} {\bibfnamefont {A.}~\bibnamefont
  {L{\'{o}}pez-D{\'{a}}valos}},\ and\ \bibinfo {author} {\bibfnamefont
  {A.}~\bibnamefont {Mart{\'{i}}n-Rodero}},\ }\bibfield  {title} {\bibinfo
  {title} {{Resonant tunneling through a small quantum dot coupled to
  superconducting leads}},\ }\href {https://doi.org/10.1103/PhysRevB.55.R6137}
  {\bibfield  {journal} {\bibinfo  {journal} {Physical Review B - Condensed
  Matter and Materials Physics}\ }\textbf {\bibinfo {volume} {55}},\ \bibinfo
  {pages} {R6137} (\bibinfo {year} {1997})}\BibitemShut {NoStop}%
\bibitem [{\citenamefont {Rogovin}\ and\ \citenamefont
  {Scalapino}(1974)}]{Rogovin1974}%
  \BibitemOpen
  \bibfield  {author} {\bibinfo {author} {\bibfnamefont {D.}~\bibnamefont
  {Rogovin}}\ and\ \bibinfo {author} {\bibfnamefont {D.~J.}\ \bibnamefont
  {Scalapino}},\ }\bibfield  {title} {\bibinfo {title} {{Fluctuation phenomena
  in tunnel junctions}},\ }\href {https://doi.org/10.1016/0003-4916(74)90430-8}
  {\bibfield  {journal} {\bibinfo  {journal} {Annals of Physics}\ }\textbf
  {\bibinfo {volume} {86}},\ \bibinfo {pages} {1} (\bibinfo {year}
  {1974})}\BibitemShut {NoStop}%
\bibitem [{\citenamefont
  {Lamic}(2025{\natexlab{b}})}]{NonEquilibriumGreenFunction}%
  \BibitemOpen
  \bibfield  {author} {\bibinfo {author} {\bibfnamefont {B.}~\bibnamefont
  {Lamic}},\ }\href {https://doi.org/10.5281/zenodo.17560811} {\bibinfo {title}
  {\text{NonEquilibriumGreenFunction.jl}: 0.2.5}} (\bibinfo {year}
  {2025}{\natexlab{b}}),\ \bibinfo {note}
  {\url{https://zenodo.org/records/17560811}}\BibitemShut {NoStop}%
\bibitem [{\citenamefont {Tinkham}(2004)}]{Tinkham2004}%
  \BibitemOpen
  \bibfield  {author} {\bibinfo {author} {\bibfnamefont {M.}~\bibnamefont
  {Tinkham}},\ }\href@noop {} {\emph {\bibinfo {title} {Introduction to
  Superconductivity}}},\ \bibinfo {edition} {2nd}\ ed.\ (\bibinfo  {publisher}
  {Dover Publications},\ \bibinfo {year} {2004})\BibitemShut {NoStop}%
\bibitem [{\citenamefont {{Levy Yeyati}}\ \emph {et~al.}(2003)\citenamefont
  {{Levy Yeyati}}, \citenamefont {Mart{\'{i}}n-Rodero},\ and\ \citenamefont
  {Vecino}}]{Yeyati2003}%
  \BibitemOpen
  \bibfield  {author} {\bibinfo {author} {\bibfnamefont {A.}~\bibnamefont
  {{Levy Yeyati}}}, \bibinfo {author} {\bibfnamefont {A.}~\bibnamefont
  {Mart{\'{i}}n-Rodero}},\ and\ \bibinfo {author} {\bibfnamefont
  {E.}~\bibnamefont {Vecino}},\ }\bibfield  {title} {\bibinfo {title}
  {{Nonequilibrium dynamics of andreev states in the kondo regime}},\ }\href
  {https://doi.org/10.1103/PhysRevLett.91.266802} {\bibfield  {journal}
  {\bibinfo  {journal} {Phys. Rev. Lett.}\ }\textbf {\bibinfo {volume} {91}},\
  \bibinfo {pages} {266802} (\bibinfo {year} {2003})}\BibitemShut {NoStop}%
\bibitem [{\citenamefont {Lamic}\ \emph {et~al.}(2020)\citenamefont {Lamic},
  \citenamefont {Meyer},\ and\ \citenamefont {Houzet}}]{BLamic2020}%
  \BibitemOpen
  \bibfield  {author} {\bibinfo {author} {\bibfnamefont {B.}~\bibnamefont
  {Lamic}}, \bibinfo {author} {\bibfnamefont {J.~S.}\ \bibnamefont {Meyer}},\
  and\ \bibinfo {author} {\bibfnamefont {M.}~\bibnamefont {Houzet}},\
  }\bibfield  {title} {\bibinfo {title} {Josephson radiation in a
  superconductor-quantum dot-superconductor junction},\ }\href
  {https://doi.org/10.1103/PhysRevResearch.2.033158} {\bibfield  {journal}
  {\bibinfo  {journal} {Phys. Rev. Res.}\ }\textbf {\bibinfo {volume} {2}},\
  \bibinfo {pages} {033158} (\bibinfo {year} {2020})}\BibitemShut {NoStop}%
\bibitem [{\citenamefont {{Press}}\ \emph {et~al.}(1992)\citenamefont
  {{Press}}, \citenamefont {{Teukolsky}}, \citenamefont {{Vetterling}},\ and\
  \citenamefont {{Flannery}}}]{numericalRecipies}%
  \BibitemOpen
  \bibfield  {author} {\bibinfo {author} {\bibfnamefont {W.~H.}\ \bibnamefont
  {{Press}}}, \bibinfo {author} {\bibfnamefont {S.~A.}\ \bibnamefont
  {{Teukolsky}}}, \bibinfo {author} {\bibfnamefont {W.~T.}\ \bibnamefont
  {{Vetterling}}},\ and\ \bibinfo {author} {\bibfnamefont {B.~P.}\ \bibnamefont
  {{Flannery}}},\ }\href@noop {} {\emph {\bibinfo {title} {{Numerical recipes
  in C. The art of scientific computing}}}},\ \bibinfo {edition} {2nd}\ ed.\
  (\bibinfo  {publisher} {Cambridge: University Press},\ \bibinfo {year}
  {1992})\BibitemShut {NoStop}%
\bibitem [{\citenamefont {Lamic}(2025{\natexlab{c}})}]{FastDysonSolver}%
  \BibitemOpen
  \bibfield  {author} {\bibinfo {author} {\bibfnamefont {B.}~\bibnamefont
  {Lamic}},\ }\href {https://doi.org/10.5281/zenodo.17643884} {\bibinfo {title}
  {\text{FastDysonSolver}}} (\bibinfo {year} {2025}{\natexlab{c}}),\ \bibinfo
  {note} {\url{https://doi.org/10.5281/zenodo.17643884}}\BibitemShut {NoStop}%
\end{thebibliography}%


\begin{thebibliography}{1}%
\makeatletter
\providecommand \@ifxundefined [1]{%
 \@ifx{#1\undefined}
}%
\providecommand \@ifnum [1]{%
 \ifnum #1\expandafter \@firstoftwo
 \else \expandafter \@secondoftwo
 \fi
}%
\providecommand \@ifx [1]{%
 \ifx #1\expandafter \@firstoftwo
 \else \expandafter \@secondoftwo
 \fi
}%
\providecommand \natexlab [1]{#1}%
\providecommand \enquote  [1]{``#1''}%
\providecommand \bibnamefont  [1]{#1}%
\providecommand \bibfnamefont [1]{#1}%
\providecommand \citenamefont [1]{#1}%
\providecommand \href@noop [0]{\@secondoftwo}%
\providecommand \href [0]{\begingroup \@sanitize@url \@href}%
\providecommand \@href[1]{\@@startlink{#1}\@@href}%
\providecommand \@@href[1]{\endgroup#1\@@endlink}%
\providecommand \@sanitize@url [0]{\catcode `\\12\catcode `\$12\catcode
  `\&12\catcode `\#12\catcode `\^12\catcode `\_12\catcode `\%12\relax}%
\providecommand \@@startlink[1]{}%
\providecommand \@@endlink[0]{}%
\providecommand \url  [0]{\begingroup\@sanitize@url \@url }%
\providecommand \@url [1]{\endgroup\@href {#1}{\urlprefix }}%
\providecommand \urlprefix  [0]{URL }%
\providecommand \Eprint [0]{\href }%
\providecommand \doibase [0]{https://doi.org/}%
\providecommand \selectlanguage [0]{\@gobble}%
\providecommand \bibinfo  [0]{\@secondoftwo}%
\providecommand \bibfield  [0]{\@secondoftwo}%
\providecommand \translation [1]{[#1]}%
\providecommand \BibitemOpen [0]{}%
\providecommand \bibitemStop [0]{}%
\providecommand \bibitemNoStop [0]{.\EOS\space}%
\providecommand \EOS [0]{\spacefactor3000\relax}%
\providecommand \BibitemShut  [1]{\csname bibitem#1\endcsname}%
\let\auto@bib@innerbib\@empty
\bibitem [{\citenamefont {Rammer}(2007)}]{Rammer2007}%
  \BibitemOpen
  \bibfield  {author} {\bibinfo {author} {\bibfnamefont {J.}~\bibnamefont
  {Rammer}},\ }\href {https://doi.org/10.1017/CBO9780511618956} {\emph
  {\bibinfo {title} {Quantum Field Theory of Non-Equilibrium States}}},\ Vol.\
  \bibinfo {volume} {9780521874}\ (\bibinfo  {publisher} {Cambridge University
  Press},\ \bibinfo {address} {Cambridge},\ \bibinfo {year} {2007})\ pp.\
  \bibinfo {pages} {1--536}\BibitemShut {NoStop}%
\end{thebibliography}%
	
\end{document}


\title{\Large Supplemental material for: \\ \normalsize Solving the Transient Dyson Equation with Quasilinear Complexity via Matrix Compression}
	\author{Baptiste LAMIC}
	\affiliation{Univ. Grenoble Alpes, CEA, IRIG-Pheliqs, F-38000 Grenoble, France}
	\date{\today}
	
	\maketitle
	
	\section{Expanded Description of the Key Equations}
	
	The objective of this section is to provide a more explicit and general formulation of the algorithm. For reader convenience, we reproduce some parts of the main text.
	
	\subsection{Problem Statement}
	We present here an algorithm to solve the Dyson equation in the transient regime within the Keldysh non-equilibrium field theory, for a given self-energy $\Sigma$. In its real-time formulation, the Dyson equation reads
	\begin{equation}
		G(t,t') = g(t,t') + \iint_{-\infty}^{\infty} g(t,t_1)\, \Sigma(t_1,t_2)\, G(t_2,t')\,dt_1\,dt_2\,.
		\label{eq:formal_dyson_equation}
	\end{equation}
	We select the Keldysh basis such that the kernels can be written as
	\begin{equation}
		G \equiv
		\begin{pmatrix}
			0 & G^{\text{A}} \\
			G^{\text{R}} & G^{\text{K}}
		\end{pmatrix}, \quad
		\Sigma \equiv
		\begin{pmatrix}
			\Sigma^{\text{K}} & \Sigma^{\text{R}} \\
			\Sigma^{\text{A}} & 0
		\end{pmatrix},
		\label{eq:basis:def}
	\end{equation}
	where the superscripts $\text{R}$, $\text{A}$, and $\text{K}$ denote the retarded, advanced, and kinetic components, as detailed in~\cite{Rammer2007}.
	Equation~\ref{eq:formal_dyson_equation} can be rewritten as
	\begin{align}
		G^{\text{R}} &= g^{\text{R}} + g^{\text{R}}\Sigma^{\text{R}}G^{\text{R}}, \\
		G^{\text{A}} &= (G^{\text{R}})^{\dagger}, \\
		G^{\text{K}} &= \left(\mathds{1} + G^{\text{R}}\Sigma^{\text{R}}\right)g^{\text{K}}\left(\mathds{1} + \Sigma^{\text{A}} G^{\text{A}}\right) + G^{\text{R}}\Sigma^{\text{K}}G^{\text{A}}, \label{eq:kinetic_dyson}
	\end{align}
	where all integrations are implicit. The orbital degrees of freedom are assumed to be discretized so that for any two kernels $(F,R)$,
	\begin{equation}
		\left[F(t,t_1)R(t_1,t')\right]_{p,q} = \sum_{k=1}^{n_o} F_{p,k}(t,t_1) R_{k,q}(t_1, t'),
	\end{equation}
	where $p,q$ label orbitals and $n_o$ is the number of orbital degrees of freedom.
	We further suppose the system is simulated from $t_0$ to $t_\text{end}$ and that the self-energy is supported only inside this domain; i.e., for all $(t,t') \notin [t_0, t_\text{end}]^2$, $\Sigma(t,t')^{R/A/K} = 0$. This restriction ensures that all integration domains lie inside $[t_0, t_\text{end}]$. The classical time stepping algorithm requires $\mathcal{O}(N^3)$ operations and $\mathcal{O}(N^2)$ space, with $N$ the number of time steps. By combining a compact discretization of the Dyson equation with matrix compression techniques, we reach $\mathcal{O}(N\log N)$ time and space complexity.
	
	\subsection{Algorithm}
	
	The algorithm relies on two main observations:
	\begin{itemize}
		\item Kernels can be efficiently compressed using hierarchical matrix formats such as HSS (Hierarchically Semi-Separable) or HODLR, greatly reducing memory and computational cost.
		\item Kernel products arising from discretized Dyson equations can be recast as matrix operations compatible with these compressed representations, preserving efficiency end-to-end.
	\end{itemize}
	\subsubsection{Equation discretization}
	The time axis is discretized into $N+1$ points on a uniform grid with step $\delta_t = (t_\text{end} - t_0)/N$. The $i$-th point is $t_i = t_0 + i\,\delta_t$. For integration, we introduce quadrature weights $w_{i,l,j}$  well-defined for $|i-j| > d$, with $d\in\mathbb{N}$, and, upon suitable redefinition, $w_{i,l,j} = 0$ whenever $l \notin \{i, \ldots, j\}$.
	We require that the weights can be factorized as
	\begin{equation}
		w_{i,l,j} = a_{i,l}\,b_{l,j} + c_{i,l,j},
	\end{equation}
	subject to the existence of integer $n_w$ such that
	\begin{equation}
		|i-j| > 2 n_w \implies
		\left\{
		\begin{array}{l}
			a_{i,j} = 1 \\
			b_{i,j} = 1 \\
			c_{i,l,j} = 0
		\end{array}
		\right. .
	\end{equation}
	This property is satisfied by any quadrature whose weights become constant far from the domain boundaries, for example, Gregory quadrature, including the trapezoidal rule with $d=0,\, n_w=1$.
	Therefore, the product of two retarded kernels $A^R$ and $B^R$ can be approximated by a Nyström discretization for $i > j + d$:
	\begin{align}
		\left[ A^R B^R \right]_{p,q}(t_i, t_j)
		&\equiv \sum_{k=1}^{n_o} \int_{t_j}^{t_i} A^R_{p,k}(t_i, t_1) B^R_{k,q}(t_1, t_j) dt_1 \\
		&\approx \delta_t \sum_{k=1}^{n_o} \sum_{l=0}^N w_{i,l,j}\, A^R_{p,k}(t_i, t_l) B^R_{k,q}(t_l, t_j),
	\end{align}
	and the entries for $i \leq j + d$ are computed by other means.  Since there are only $\mathcal{O}(N)$ such elements, such elements and they are located near the diagonal, all of these entries can be evaluated in $\mathcal{O}(N)$ time.  Inserting the quadrature decomposition, we almost rewrite the kernel product as a matrix product
	\begin{align}
		\left[A^R B^R\right]_{p,q}(t_i, t_j) \approx
		&\delta_t \sum_{k=1}^{n_o} \sum_{l=0}^N a_{i,l}\,A^R_{p,k}(t_i, t_l)\,b_{l,j} B^R_{k,q}(t_l, t_j) \\
		&+ \delta_t \sum_{k=1}^{n_o} \sum_{l=j}^i c_{i,l,j} A^R_{p,k}(t_i, t_l) B^R_{k,q}(t_l, t_j).
		\label{eq:first_factorisation}
	\end{align}
	This suggests defining the correction term   
	\begin{equation}
		\mathcal{C}[A^R, B^R]_{p,q}(t_i, t_j) \equiv \delta_t \sum_{k=1}^{n_o} \sum_{l = j}^{i} c_{i,l,j} A^R_{p,k}(t_i, t_l) B^R_{k,q}(t_l, t_j) +  \Pi_d(i-j) \left[A^R B^R\right]_{p,q}(t_i, t_j),
	\end{equation}
	where $\Pi_d(i-j) = 1$ if $|i-j| \le d$ and 0 otherwise. Only $\mathcal{O}(N)$ coefficients $c_{i,l,j}$, all located close to the diagonal, are nonzero, so $\mathcal{C}$ can be computed in $\mathcal{O}(N)$ time.
	Thus for all $i, j \in \{1, \ldots, N\}$, the kernel product can be expressed as
	\begin{equation}
		\left[ A^R B^R \right]_{p,q}(t_i, t_j) \approx \delta_t \sum_{k=1}^{n_o} \sum_{l=0}^N a_{i,l} A^R_{p,k}(t_i, t_l)\, b_{l,j} B^R_{k,q}(t_l, t_j) + \mathcal{C}[A^R, B^R]_{p,q}(t_i, t_j).
	\end{equation}
	Equipped with this expression, we can reformulate the kernel product as a matrix product. 
	\subsubsection{Matrix Notation.}
	Let $\sigma: (p,i) \mapsto p + i\,n_o$ be the map from orbital and time indices to matrix indices. we define $\mathbf{A} \in \mathbb{C}^{n_o N \times n_o N}$ as
	\begin{equation}
		\mathbf{A}_{\sigma(p,i),\, \sigma(q,j)} \equiv A_{p,q}(t_i, t_j).
	\end{equation}
	We define the matrix form of $a$ and $b$ as
	\begin{align}
		\mathbf{a}_{\sigma(p,i), \sigma(q,j)} &= a_{i,j}, \\
		\mathbf{b}_{\sigma(p,i), \sigma(q,j)} &= b_{i,j}.
	\end{align}
	And finally, we introduce the blockwise Hadamard product $\odot$ defined as
	\begin{equation}
		\left[\mathbf{A} \odot \mathbf{B}\right]_{\sigma(p,i),\,\sigma(q,j)} \equiv \sum_{k=1}^{n_o} \mathbf{A}_{\sigma(p,i),\,\sigma(k,j)}\, \mathbf{B}_{\sigma(k,i),\,\sigma(q,j)}.
	\end{equation}
	Combining all our ingredients, we obtain the compact expression 
	\begin{equation}
		[\mathbf{A}^R \mathbf{B}^R] = (\mathbf{a} \odot \mathbf{A}^R)\,(\mathbf{b} \odot \mathbf{B}^R) + \mathbf{C}[A^R, B^R].
		\label{eq:kernel_product_as_matrix_product}
	\end{equation}
	The matrix $\mathbf{C}[A^R, B^R]$ is extremely sparse and can be evaluated cheaply using standard methods in $\mathcal{O}(N)$ operations, which is trivial when using trapezoidal quadrature. The blockwise Hadamard product is computed efficiently by exploiting that $a_{i,j}$ and $b_{i,j}$ deviate from unity only near the diagonal. Since matrices $\mathbf{A}$ and $\mathbf{B}$ are already in HSS format, the evaluation of \cref{eq:kernel_product_as_matrix_product} achieves overall $\mathcal{O}(N)$ complexity. We can now turn to solving the Dyson equation.
	
	\subsubsection{Application to the Dyson Equation.}
	Let's use this discretization of the Kernel product to discretize the retarded Dyson equation. 
	First we define $F^R \equiv g^R \Sigma^R$,
	\begin{equation}
		\mathbf{G}^R = \mathbf{g}^R + (\mathbf{a} \odot \mathbf{F}^R)\, (\mathbf{b} \odot \mathbf{G}^R) + \mathbf{C}[F^R, G^R].
	\end{equation}
	We regroup the terms,
	\begin{equation}
		(\mathbf{b} \odot \mathbf{G}^R) = \mathbf{g}^R + (\mathbf{a} \odot \mathbf{F}^R)\,(\mathbf{b} \odot \mathbf{G}^R) + \mathbf{C}[F^R, G^R] - ((\mathbf{1} - \mathbf{b}) \odot \mathbf{G}^R), 
	\end{equation}
	and define
	\begin{equation}
		\mathbf{S} \equiv \mathbf{C}[F^R, G^R] - ((\mathbf{1} - \mathbf{b}) \odot \mathbf{G}^R).
	\end{equation}
	By design, $\mathbf{S}$ is only nonzero near the diagonal, i.e., $\mathbf{S}_{\sigma(p,i),\,\sigma(q,j)} \ne 0 \implies |i-j| \le \max(2n_w, d)$. Consequently, $\mathbf{S}$ is efficiently computable in $\mathcal{O}(N)$ time.
	The Dyson equation thus becomes
	\begin{equation}
		(\mathbf{b} \odot \mathbf{G}^R) = [\mathrm{Id} - (\mathbf{a} \odot \mathbf{F}^R)]^{-1} \left(\mathbf{g}^R + \mathbf{S}\right),
	\end{equation}
	with $\mathrm{Id}$ the identity. This equation can be solved efficiently in compressed format, yielding the retarded Green function throughout most of the domain and enabling evaluation of the full Green function over the complete domain in $\mathcal{O}(N)$ operations when starting from already compressed matrices. As discussed in the main text, the compression bottleneck is bypassed using stochastic compression methods, which preserves the quasi-linearity of the entire approach. 
	For trapezoidal quadrature, these expressions reduce to those presented in the main text.

	\bibliography{biblio}